\documentclass[journal,twoside]{IEEEtran}

\usepackage{amsmath}
\usepackage{amssymb}
\usepackage{bm}
\usepackage{graphicx}
\usepackage{cite}
\usepackage{url}
\usepackage{stfloats}

\graphicspath{{plot/}}

\newcommand{\bh}{\ensuremath{\bm{h}}}
\newcommand{\bv}{\ensuremath{\bm{v}}}
\newcommand{\varn}{\ensuremath{\sigma_{n}^{2}}}
\newcommand{\varw}{\ensuremath{\sigma_{w}^{2}}}
\newcommand{\varz}{\ensuremath{\sigma_{z}^{2}}}

\newcommand{\rvq}{\ensuremath{\mathrm{rvq}}}

\newcommand{\bvhh}{\ensuremath{\bm{v}(\hat{\bm{h}})}}
\newcommand{\bhh}{\ensuremath{\hat{\bm{h}}}}
\newcommand{\bI}{\ensuremath{\bm{I}}}
\newcommand{\bn}{\ensuremath{\bm{n}}}
\newcommand{\bw}{\ensuremath{\bm{w}}}
\newcommand{\bvT}{\ensuremath{\bm{v}_{T}}}
\newcommand{\br}{\ensuremath{\bm{r}}}
\newcommand{\bBT}{\ensuremath{\bm{B}_{T}}}
\newcommand{\bVT}{\ensuremath{\bm{V}_{T}}}
\newcommand{\Lb}{\ensuremath{\bar{L}}}
\newcommand{\T}{\ensuremath{\bar{T}}}
\newcommand{\D}{\ensuremath{\bar{D}}}
\newcommand{\B}{\ensuremath{\bar{B}}}

\newcommand{\diff}{\ensuremath{\mathrm{d}}}
\newcommand{\dNt}{\ensuremath{d({N_t)}}}
\newcommand{\cNt}{\ensuremath{c({N_t})}}
\newcommand{\bC}{\ensuremath{\bm{C}}}
\newcommand{\bR}{\ensuremath{\bm{R}}}
\newcommand{\mC}{\ensuremath{\mathcal{C}}}
\newcommand{\N}{\bar{N}_r}
\newcommand{\bH}{\ensuremath{\bm{H}}}
\newcommand{\bHh}{\ensuremath{\hat{\bm{H}}}}
\newcommand{\bvH}{\ensuremath{\bm{v}(\bm{H})}}
\newcommand{\bvHh}{\ensuremath{\bm{v}(\hat{\bm{H}})}}
\newcommand{\srvq}{\gamma_{\rvq}}
\newcommand{\var}{\ensuremath{\mathrm{var}}}
\newcommand{\rNt}{\ensuremath{r({N_t})}}
\newcommand{\sNt}{\ensuremath{s({N_t})}}
\newcommand{\Tu}{\ensuremath{\T^o_u}}
\newcommand{\Bu}{\ensuremath{\B^o_u}}
\newcommand{\Du}{\ensuremath{\D^o_u}}
\newcommand{\mL}{\ensuremath{\mathcal{L}}}
\newcommand{\me}{\mathrm{e}}
\newcommand{\zb}{\zeta_{\B}}

\newcounter{mytempeqncnt}

\newtheorem{theorem}{Theorem}
\newtheorem{lemma}{Lemma}

\begin{document}

\title{Optimization of Training and Feedback Overhead for Beamforming
  over Block Fading Channels}

\author{Wiroonsak Santipach,~\IEEEmembership{Member,~IEEE,} 
        and Michael L. Honig,~\IEEEmembership{Fellow,~IEEE}%
\thanks{This work was supported by the U.S. Army Research Office under
grant W911NF-07-1-0028 and the National Science Foundation under grant
CCR-0310809, and was presented in part at the IEEE International
Symposium on Information Theory, Seattle, WA, July 2006, and the IEEE
Wireless Communications and Networking Conference, Hong Kong, China,
March, 2007.}%
\thanks{W. Santipach is with the Department of Electrical Engineering,
Faculty of Engineering, Kasetsart University, Bangkok 10900, Thailand;
Email: wiroonsak.s@ku.ac.th.}%
\thanks{M. L. Honig is with the Department of Electrical Engineering and
Computer Science, Northwestern University, Evanston, Illinois 60208 USA;
Email: mh@eecs.northwestern.edu.}%
\thanks{Communicated by G. Taricco, Associate Editor for Communications.}}

% The paper headers 

\markboth{IEEE Transactions on Information Theory}{Santipach and
  Honig: Optimization of Training and Feedback Overhead for
  Beamforming over Block Fading Channels }

\maketitle

\begin{abstract}
We examine the capacity of beamforming over a single-user,
multi-antenna link taking into account the overhead due to channel
estimation and limited feedback of channel state information.
Multi-input single-output (MISO) and multi-input multi-output (MIMO)
channels are considered subject to block Rayleigh fading.  Each
coherence block contains $L$ symbols, and is spanned by $T$ training
symbols, $B$ feedback bits, and the data symbols. The training symbols
are used to obtain a Minimum Mean Squared Error estimate of the
channel matrix.  Given this estimate, the receiver selects a transmit
beamforming vector from a codebook containing $2^B$ {\em i.i.d.}
random vectors, and sends the corresponding $B$ bits back to the
transmitter. We derive bounds on the beamforming capacity for MISO and
MIMO channels and characterize the optimal (rate-maximizing) training
and feedback overhead ($T$ and $B$) as $L$ and the number of transmit
antennas $N_t$ both become large.  The optimal $N_t$ is limited by the
coherence time, and increases as $L/\log L$.  For the MISO channel the
optimal $T/L$ and $B/L$ (fractional overhead due to training and
feedback) are asymptotically the same, and tend to zero at the rate
$1/\log N_t$.  For the MIMO channel the optimal feedback overhead
$B/L$ tends to zero faster (as $1/\log^2 N_t$).
\end{abstract}


%%%%%%%%%%%%%%%%%%%%%%%%%%%%%%%%%%%%%%%%%%%%%%%%%%%%%%%%%%%%%%%%%%%%%%
\section{Introduction}
%%%%%%%%%%%%%%%%%%%%%%%%%%%%%%%%%%%%%%%%%%%%%%%%%%%%%%%%%%%%%%%%%%%%%%

\IEEEPARstart{W}{ith} perfect channel knowledge at the transmitter and
receiver, the capacity of a multi-antenna system with independent
Rayleigh fading increases with the number of antennas
\cite{telatar99,foschini98}.  In practice, the channel estimate at the
receiver will not be perfect, and furthermore, this estimate must be
quantized before it is relayed back to the transmitter.  This has
motivated work on the performance of feedback schemes with imperfect
channel knowledge
\cite{visotsky01,zhou02,simon03,YooGol06,skoglund07,blum03,lau04}, and
the design and performance of limited feedback schemes for Multi-Input
Multi-Output (MIMO) and Multi-Input Single-Output (MISO) channels
(e.g., see
\cite{narula98,love03,mukkavilli03,roh_it04,lau04,SanHon09,dai_liu,
  chun_love,love06} and the recent survey paper \cite{LovHeaLau08}).
All of the previous work on limited feedback assumes perfect channel
knowledge at the receiver.  Here we consider a model that takes into
account {\em both} imperfect channel estimation at the receiver {\em
  and} limited channel state feedback.

We focus on single-user MISO and MIMO links with rank-one precoders
(beamforming), and study the achievable rate as a function of overhead
for channel estimation and channel state feedback. Our objective is to
characterize the optimal amount of overhead and the associated
achievable rate, and to show how those scale with the system size
(i.e., as the number of transmit and/or receive antennas become
large).  Motivated by practical systems, a pilot-based scheme for
channel estimation is assumed.  Given a finite coherence time, the
number of antennas that can be used effectively is limited by the
channel estimation error and quantization error associated with the
transmit beam. We show how the optimal (rate-maximizing) number of
transmit antennas scales with the system size.

More specifically, an independent identically distributed ({\em i.i.d.})
block Rayleigh fading channel is considered in which 
the channel parameters are stationary within each coherence block, 
and are independent from block to block. 
The block length $L$ is assumed be constant, and the transmitted
codewords span many blocks, so that the maximum achievable rate is the
ergodic capacity.  Each coherence block contains $T$ training symbols
and $D$ data symbols.  Furthermore, we assume that after transmission
of the training symbols, the transmitter waits for the receiver to
relay $B$ bits over a feedback channel, which specify a particular
beamforming vector.  This delay, in addition to the $T$ training
symbols, must occur within the coherence block, and is therefore 
counted as part of the packet overhead.\footnote{An implicit 
assumption is that the transmitter cannot learn the channel 
by detecting a received signal in the reverse direction, as in some
Time-Division Duplex systems (e.g., see \cite{SteSab08}). Although
the feedback overhead is counted as part of the coherence time, 
a similar penalty arises with a 
Frequency-Division Duplex model \cite{XieGeo06}.}

We assume that the receiver computes a Minimum Mean Square Error
(MMSE) estimate of the channel, based on the training symbols, and
uses the noisy channel estimate to choose a transmit beamforming
vector. The Random Vector Quantization (RVQ) scheme in
\cite{SanHon05,chun_love,SanHon09} is assumed
in which the beamformer is selected
from a codebook consisting of $2^B$ random vectors, which are
independent and isotropically distributed, and known {\em a priori} at
the transmitter and receiver.  The associated codebook index is
relayed using $B$ bits via a noiseless feedback channel to the
transmitter.  The capacity of this scheme with perfect channel
estimation is analyzed in \cite{SanHon05,love06,chun_love,Jindal06,SanHon09}.  
It is shown in \cite{SanHon09} that the RVQ codebook is optimal 
(i.e., maximizes the capacity)
in the large system limit in which number of transmit antennas $N_t$
and $B$ tend to infinity with fixed ratio $\B = B/N_t$.  In
\cite{commag04,SanHon09}, RVQ has been observed to give essentially
optimal performance for systems with small $N_t$.  
Furthermore, for the MISO channel
the performance averaged over the random codebooks can be
explicitly computed \cite{chun_love}.

The capacity with MMSE channel estimates at the receiver (with or
without limited feedback) is unknown.\footnote{An analysis of the
  error rate for MIMO links with MMSE channel estimates without
  feedback is given in \cite{taricco05,taricco10}.}  We derive upper
and lower bounds on the capacity with RVQ and limited feedback, which
are functions of the number of training symbols $T$ and feedback bits
$B$.  Given a fixed block size, or coherence time $L$, we then
optimize the capacity bounds over $B$ and $T$.  Namely, small $T$
leads to a poor channel estimate, which decreases capacity, whereas
large $T$ leads to an accurate channel estimate, but leaves few
symbols in the packet for transmitting the message.  This trade-off
has been studied in \cite{hassibi,sun_globecom} for MIMO channels
without feedback.  Here there is also an optimal amount of feedback
$B$, which increases with the training interval $T$.  That is, more
feedback is needed to quantize more accurate channel estimates.

We characterize the optimal overhead due to training and feedback
in the large system limit as the coherence time $L$
and number of transmit antennas $N_t$ both tend to infinity
with fixed ratio $\bar{L} = L/N_t$. For the MIMO channel we
also let the number of receiver antennas $N_r \to \infty$
with fixed $N_t / N_r$. This allows a characterization of
the achievable rate as a function of the number of feedback
bits per degree of freedom \cite{SanHon09}.\footnote{See also 
the tutorial on large random matrix theory \cite{TulVer04FT}.}

For both MISO and MIMO channels the optimal normalized training 
$\T=T/L$, which maximizes the bounds on capacity, tends to zero at the
rate $1/\log N_t$. For the MISO channel the normalized feedback $\B= B/L$
also tends to zero at this rate. Moreover, the training and feedback
require the same asymptotic overhead. For the MIMO channel
the optimal $\B=B/L$ tends to zero at the rate $1/\log^2 N_t$.
Hence the overhead due to feedback is lower for the MIMO channel
than for the MISO channel. This is apparently due to the
additional degrees of freedom at the receiver, which
can compensate for the performance loss associated with quantization error.

For both MISO and MIMO channels, the optimal $T$ increases as $N_t /
\log N_t$, and we observe that the associated capacity can be achieved
by activating only $N_t / \log N_t$ antennas (assuming $N_t$ increases
linearly with $L$).  Equivalently, for this pilot-based scheme with
limited feedback, the optimal number of (active) transmit antennas
increases as $L/ \log L$. Hence the training and feedback overhead
pose a fundamental limit on the number of antennas that can be
effectively used. The capacity with optimized overhead grows as $\log
N_t$. This is the same as with perfect channel knowledge; however,
there is a second-order loss term, which increases as $\log \log N_t$.

A similar type of model for optimizing feedback overhead has been
previously considered in \cite{XieGeo06}. A key difference is that
here the relation between training and channel estimation error is
explicitly taken into account.  The model we present is also closely
related to the two-way limited feedback system considered in
\cite{LovAuY08,LovAuY09} (see also \cite{SteSab08}).  However, here
the feedback channel is simply modeled with a fixed rate (i.e., is not
the result of an optimization), and reflects the likelihood that the
forward channel may be quite different from the reverse (feedback)
channel. Also, the scaling of the optimal overhead and capacity with
system size, given a fixed coherence time and fixed feedback rate, is
not addressed in the preceding references.  Similar types of overhead
and capacity scaling results to those presented here are presented in
\cite{AgaHon05} for a single-user wideband multi-carrier channel and
in \cite{CheBerHon08} for the cellular downlink based on Orthogonal
Frequency Division Multiple Access.

The rest of the paper is organized as follows. Section~\ref{sys_mod}
describes the multi-antenna channel model. Bounds on the beamforming
capacity for the MISO channel with channel estimation and limited
feedback are presented in Section ~\ref{tmiso} along with a
characterization of the optimal (capacity-maximizing) training and
feedback lengths in the large system limit.  Corresponding results for
the MIMO channel are presented in Section~\ref{tmimo}.  Numerical
results for finite-size MISO and MIMO channels are shown in
Section~\ref{numerical}, and conclusions are presented in
Section~\ref{conclude}.

%%%%%%%%%%%%%%%%%%%%%%%%%%%%%%%%%%%%%%%%%%%%%%%%%%%%%%%%%%%%%%%%%%%%%
\section{System Model}
\label{sys_mod}
%%%%%%%%%%%%%%%%%%%%%%%%%%%%%%%%%%%%%%%%%%%%%%%%%%%%%%%%%%%%%%%%%%%%%

We consider a point-to-point {\em i.i.d.} block fading
channel with $N_t$ transmit antennas and $N_r$ receive antennas.
A rich scattering environment is assumed so that the channel gains
corresponding to different pairs of transmit/receive antennas are
independent and Rayleigh distributed.
The $i$th $N_r \times 1$ received vector in a
particular block is given by
\begin{equation}
 \br (i) = \bH \bv b (i) + \bn (i) \qquad \text{for} \quad
1 \le i \le D
\label{eq:model}
\end{equation}
where $\bH$ is an $N_r \times N_t$ channel matrix whose elements are
independent, complex Gaussian random variables with zero mean and unit
variance, $\bv$ is an $N_t \times 1$ unit-norm beamforming vector, $b$
is the transmitted symbol with unit variance, $\bn$ is additive white
Gaussian noise (AWGN) with covariance $\varn \bI$, and $D$ is the
number of data (information) symbols in a block.

\subsection{Random Vector Quantization}
In prior work \cite{SanHon09},
we have analyzed the channel capacity with perfect 
channel knowledge at the receiver, but with {\em limited} 
channel knowledge at the transmitter.
Specifically, the optimal beamformer is quantized
at the receiver, and the quantized version is relayed
back to the transmitter. Given the quantization codebook
$\mathcal{V} = \{ \bv_1, \ldots, \bv_{2^B} \}$, which is
also known {\em a priori} at the transmitter,
and the channel $\bH$,
the receiver selects the quantized beamforming vector
to maximize the instantaneous rate,
\begin{equation}
  \bvH = \arg \max_{\bv_j \in \mathcal{V}} \left\{ \log (
  1 + \rho \| \bH \bv_j \|^2 ) \right\}
\label{vH}
\end{equation}
where $\rho = 1/ \varn$ is the background signal-to-noise ratio (SNR).
The (uncoded) index for the rate-maximizing beamforming vector is
relayed to the transmitter via an error-free feedback link.  The
capacity depends on the beamforming codebook $\mathcal{V}$ and
$B$. With unlimited feedback ($B \to \infty$) the $\bvH$ that
maximizes the capacity is the properly normalized eigenvector of
$\bH^{\dag} \bH$, which corresponds to the maximum eigenvalue.

We will assume that the codebook vectors are
independent and isotropically distributed over the unit sphere.
It is shown in \cite{SanHon05,SanHon09}
that this RVQ scheme is optimal 
(i.e., maximizes the achievable rate) in the
large system limit in which $(B, N_t, N_r) \to \infty$
with fixed normalized feedback $\B = B / N_t$ and $\N = N_r / N_t$.
(For the MISO channel $N_r=1$.)
Furthermore, the corresponding capacity grows as $\log (\rho N_t )$,
which is the same order-growth as with perfect
channel knowledge at the transmitter.
Although strictly speaking, RVQ is suboptimal for a
finite-size system, numerical results indicate that the average performance
is often indistiguishable from 
the performance with optimized codebooks \cite{commag04,SanHon09}.

\subsection{Channel Estimation}
In addition to limited channel information at the transmitter, here we
also account for channel estimation error at the receiver.
Letting $\bHh$ be the estimated channel matrix, the
receiver selects $\bvHh$ assuming that $\bHh$ is the
actual channel, i.e.,
\begin{equation}
  \bvHh = \arg \max_{\bv_j \in \mathcal{V}} \left\{ \log (
  1 + \rho \| \bHh \bv_j \|^2 ) \right\} .
\label{bvhh}
\end{equation}

We will assume that the receiver computes the linear MMSE estimate
of $\bH$ given the received vectors corresponding to $T$ training vectors.
Specifically, the transmitter transmits $T$ training symbols
$b_T (1) , \cdots , b_T (T)$, where the training symbol
$b_T (i)$ modulates the corresponding beamforming vector $\bvT (i) $.
For the MISO channel the row vector of $T$ received samples is given by
\begin{equation}
  \br_T = \bh \bVT \bBT + \bn_T
\end{equation}
where the channel $\bh$ is a $1 \times N_t$ row vector,
$\bVT = [ \bvT(1) \cdots \bvT(T)]$,
$\bBT = \mathrm{diag}\{ b_{T} (i) \}$,
and $\bn_T = [n(1) \cdots n(T)]$.
The channel estimate is $\hat{\bh} = \br_T \bC$, where
the $T \times N_t$ linear MMSE channel estimation filter is given by
\begin{eqnarray}
  \bC & = & \arg \min_{\tilde{\bC}}
      E [ \| \bh - \br_T \tilde{\bC} \|^2 ]\\ & = &
      \bVT \bBT (\bVT^{\dag}\bVT + \varn \bI)^{-1}. \label{bc}
\end{eqnarray}
The MSE
\begin{equation}
  \varw = E[ \| h_i - \hat{h}_i \|^2 ] =
      1 - \frac{1}{N_t} \text{trace} \{ \bC^\dag \bR_T \bC \}
\end{equation}
where $h_i$ and $\hat{h}_i$ are $i$th elements of
$\bh$ and $\hat{\bh}$, respectively, and the
received covariance matrix
\begin{equation}
  \bR_T = E[ \br_T^\dag \br_T ]= \bBT \bVT^{\dag} \bVT
  \bBT^{\dag} + \varn \bI .
\end{equation}

The preceding expressions also apply to the MIMO channel
where the estimation is for a particular row of $\bH$.
That is, $\bC$ is replaced by $\bC_i$, which is
applied to the $i$th receiver antenna, and used to estimate
the $i$th row of $\bH$. The MSE for each element of
$\bH$ therefore remains the same.

Because the elements of $\bH$ are assumed to be complex
{\em i.i.d.} Gaussian random variables, we have
\begin{equation}
  \bH  = \bHh + \bw
\label{bhh}
\end{equation}
where the estimate $\bHh$ and the error matrix $\bw$ are independent,
and each contain {\em i.i.d.} complex Gaussian elements.
The elements of $\bw$ have zero mean and variance $\varw$, so that
$\bHh$ has zero mean and covariance $(1 - \varw) \bI$.

The variance $\varw$ clearly decreases as $T$ increases.
Furthermore, since the beamforming vectors during training $\bVT$ are
known {\em a priori} to the transmitter and receiver,
those can be chosen to minimize the MSE.
It is shown in \cite{hassibi} that the corresponding set of (unit-norm)
beamforming vectors achieves the Welch bound with equality.
We therefore have that \cite{rupf94}
\begin{alignat}{2}
  \bVT \bVT^{\dag} &= \T \bI \quad &&\text{if} \quad T > N_t, \\
  \bVT^{\dag} \bVT &= \bI &&\text{if} \quad T \le N_t . \label{vc}
\end{alignat}
Applying \eqref{bc}-\eqref{vc}, we obtain the
variance of the estimation error
\begin{equation}
  \varw = \left\{ \begin{array}{l@{,\quad}l}
         1 - \frac{\T}{1 + \rho^{-1}}& \T < 1\\
         \frac{1}{1 + \rho \T} & \T \ge 1 \end{array} \right. .
\label{vw}
\end{equation}

\subsection{Ergodic Capacity}
In what follows, we assume that the forward and feedback links are
time-division multiplexed, and each block consists of $T$ training
symbols, $B$ feedback bits, and $D$ data symbols.  Given that the size
of each block is $L$ symbols, we have the constraint
\begin{equation}
  L = T + \mu B + D
\label{contr}
\end{equation}
where $\mu$ is a conversion factor, which relates bits to symbols.
Our objective is to maximize the ergodic capacity, which is the
maximum mutual information between $b$ and $\br$,
\begin{equation}
  \max_{T,B} \ \{ C = E [ \max_{p_b} I (\br; b | \bH, \bHh, \bvHh) ] \}
\label{CE}
\end{equation}
subject to \eqref{contr}, where $p_b$ is the probability density
function (pdf) for the transmitted symbol $b$, and
the expectation is over the channel $\bH$, the estimation error
$\bw$, and the RVQ codebook $\mathcal{V}$.
Determining the ergodic capacity of RVQ with
channel estimation appears to be
intractable, so instead we derive upper and lower bounds, which are
functions of $D$, $B$, and $T$.  We then maximize both bounds
over $\{D, B, T\}$, subject to \eqref{contr}.

% Double-column Eq -----------------------------------------------
\begin{figure*}[!b]
  \normalsize
  \setcounter{mytempeqncnt}{\value{equation}}
  \setcounter{equation}{27} 
  \vspace*{4pt}
  \hrulefill
  \begin{equation} 
  \label{dNt}
  \dNt = \frac{1}{2} \sqrt{ \frac{1}{N_t} + \left( 1 + \frac{1}{N_t}
    \right) \frac{\Gamma \left(1 + \frac{2}{N_t-1} \right) - \Gamma^2
      \left(1 + \frac{1}{N_t-1} \right)(1 + 2^{-\B N_t
      })^{-\frac{2}{N_t-1}}} {\left( 2^{\B + \frac{\B}{N_t-1}} -
      \Gamma \left(1 + \frac{1}{N_t-1} \right) \right)^2}}
\end{equation}
\setcounter{equation}{\value{mytempeqncnt}}
\end{figure*}
%------------------------------------------------------------------

%%%%%%%%%%%%%%%%%%%%%%%%%%%%%%%%%%%%%%%%%%%%%%%%%%%%%%%%%%%%%%%%%%%
\section{Multi-Input Single-Output Channel}
\label{tmiso}
%%%%%%%%%%%%%%%%%%%%%%%%%%%%%%%%%%%%%%%%%%%%%%%%%%%%%%%%%%%%%%%%%%%

\subsection{Capacity Bounds}
We first consider a MISO channel with $1 \times N_t$ channel vector
$\bh$.  Applying Jensen's inequality, we obtain the upper bound on
ergodic capacity
\begin{align}
  C & = E [ \max_{p_b} I ( b; \br | \bhh, \bvhh, \bh ) ] \\
    & =  E [ \log ( 1 + \rho | \bh \bvhh |^2) ] \label{lb}\\
    & \le  \log ( 1 + \rho E [ | \bh \bvhh |^2 ]) \label{js}
\end{align}
where the maximizing pdf is Gaussian, and the expectation is over
$\bh$, the estimation error $\bw$, and the random codebook
$\mathcal{V}$.  Substituting $\bh = \bhh + \bw$ into the expectation
in \eqref{js} and simplifying gives
\begin{equation}
  E [ | \bh \bvhh |^2 ]  =  \varw + E [ |\bhh \bvhh |^2 ] .
\label{evh}
\end{equation}
Since $\| \bhh \|^2$ and $\nu \triangleq | \bhh \bvhh |^2 / \| \bhh
\|^2$ are independent \cite{chun_love,roh_it04}, we have
\begin{equation}
   E [ | \bhh \bvhh |^2 ] = E [\| \bhh \|^2 ] E [\nu ] = (1 -
   \varw) N_t E [ \nu ] .
\end{equation}

With RVQ we have
\begin{equation}
  \nu = \max_{1 \le j \le 2^B} \{ \nu_j = | \bhh \bv_j |^2
  / \| \bhh \|^2 \}
\end{equation}
where the $\nu_j$'s are {\em i.i.d.} with pdf given in \cite{mukkavilli03}.
The pdf for $\nu$ and associated
mean can be explicitly computed \cite{chun_love}.
The mean is given by
\begin{equation}
  E [ \nu ] = 1 - 2^B B \left( 2^B, \frac{N_t}{N_t - 1} \right)
\label{mean}
\end{equation}
where the beta function $B(m,n) = \int_0^1 t^{m-1} (1 - t)^{n-1} \,
\diff t$ for $m$ and $n > 0$.
We can bound $E[ \nu ]$ as follows.
\begin{lemma}
\label{enu}
For $\B \ge 0$ and $N_t \ge 2$,
\begin{eqnarray}
  E [ \nu ] & \le & 1 - 2^{-\B} + \frac{1 + (\gamma - 1) 2^{-\B} +
  2^{-\B N_t}}{N_t -1} \label{12b1}\\
  E [ \nu ] & \ge & 1 - 2^{-\B}
  \label{12b2}
\end{eqnarray}
where $\gamma = 0.5772 \ldots$ is the Euler constant.
\end{lemma}
The proof is given in Appendix~\ref{snr_bnds}.  We note that $E [ \nu
] \to 1 - 2^{-\B}$ as $N_t \to \infty$.  Substituting
\eqref{evh}-\eqref{12b1} into \eqref{js} gives an upper bound on
capacity.

To derive a lower bound on capacity, we use
the estimation error equation $\bh = \bhh + \bw$ to write
\begin{equation}
  r(i) = (\bhh \bvhh) b(i) + \underbrace{ (\bw \bvhh)
  b(i) + n(i) }_{z(i)} .
\end{equation}
Since $\bw$ and $\bhh$ are independent, it follows that $ E [
z(i) b(i) ] = 0$.  It is shown in \cite{hassibi,medard00} that
replacing $z(i)$ with a zero-mean Gaussian random variable minimizes
the mutual information $I ( r; b | \bhh, \bvhh)$ and therefore gives a
lower bound on the capacity with channel estimation and quantized
beamforming.  The lower bound is maximized when $b(i)$ has a Gaussian
pdf, i.e.,
\begin{align}
  C  &\ge  E [ \max_{p_b} \min_{p_z} I ( r;b | \bhh, \bvhh) ]\\
     &=   E \left[ \log \left(1 + \frac{| \bhh \bvhh
    |^2}{\varz} \right) \right] \label{ell}
\end{align}
where $p_z $ and $\varz$ denote the pdf and variance for $z$,
respectively.  We derive the following lower bound on $C$ by applying
the inequality in \cite{bental}.
\begin{lemma}
\label{l_bental}
\begin{multline}
  E \left[ \log \left(1 + \frac{1}{\varz} | \bhh \bvhh
    |^2 \right) \right] \\
  \ge
  ( 1 - \dNt )\log
  \left( 1 +  \frac{1}{\varz} E [  | \bhh \bvhh
    |^2 ] \right)
\label{l2}
\end{multline}
where $\dNt$ is shown in \eqref{dNt} and
the gamma function $\Gamma(m) = \int_0^{\infty} t^{m-1} \me^{-t} \,
\diff t$ for $m > 0$.
\addtocounter{equation}{1}
\end{lemma}
The proof is given in Appendix \ref{pf_bental}.
We note that $\dNt \to 0$ as $N_t \to \infty$.

To obtain a lower bound on capacity $C$, we substitute $\varz =
\varw + \varn$, \eqref{12b2}, and \eqref{l2}-\eqref{dNt} into
\eqref{ell}.  The capacity bounds are summarized as follows.
\begin{theorem}
\label{bnd}
  The capacity for a MISO channel with channel estimation variance
  $\varw$ and normalized feedback $\B$ satisfies
  \begin{equation}
    C_l \le C \le C_u \quad \text{for} \quad \B \ge 0 \ \text{and} \
    N_t \ge 2
  \end{equation}
where
\begin{align}
  C_l &=  (1 - \dNt) \log \left(1 + \rho \frac{1 - \varw}{1 + \rho
  \varw} (1 - 2^{-\B}) N_t \right), \label{eq:Cl}\\
  C_u &= \log \bigg( 1 + \rho \varw + \rho (1 - \varw)N_t \nonumber\\
      &\quad \times 
      \bigg( 1 - 2^{-\B} + \frac{1 + (\gamma - 1)
      2^{-\B} + 2^{-\B N_t}}{N_t-1} \bigg) \bigg).
\end{align}
\end{theorem}
The gap between the two bounds tends to zero as
$\rho \to 0$ (since both $C_u$ and $C_l$ tend to zero),
%$\varw$
and as $N_t \to \infty$.
With fixed $\B$ and $\varw$ the bounds (and the capacity)
grow as $O(\log N_t)$ as $N_t \to \infty$.
Substituting \eqref{vw} for $\varw$ gives the bounds
as a function of training $T$.

\begin{figure}
  \centering
  \includegraphics[width=3.25in]{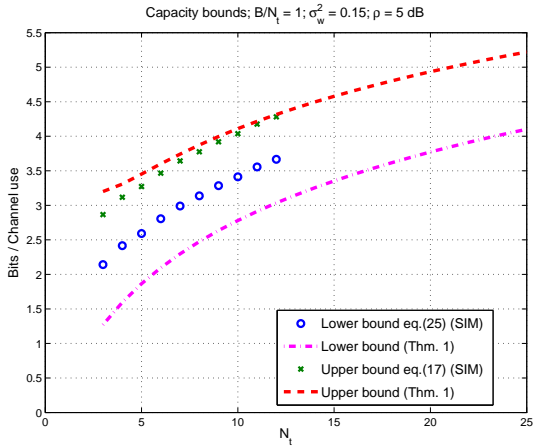} % For 2-column, use 3in
  \caption{The capacity bounds in Theorem~\ref{bnd} (bits/channel use)
  versus number of transmit antennas.}
  \label{bnds}
\end{figure}

Fig.~\ref{bnds} compares the bounds in Theorem~\ref{bnd}
with \eqref{lb} and the tighter lower bound \eqref{ell}. 
The bounds are plotted versus $N_t$ with
parameters $B/N_t =1$ (one bit per antenna coefficient),
$\sigma_w^2 =0.15$, and SNR $\rho= 5$ dB.
The tighter bounds, which are analytically intractable,
are evaluated by Monte Carlo simulation and shown as $\circ$'s and
$\times$'s in the figure.  The plots show that the upper bound in
Theorem~\ref{bnd} is close to \eqref{lb} even for small $N_t$ while
the lower bound in the Theorem is close to \eqref{ell} for much larger
$N_t$.  Since RVQ requires an exhaustive search over
the codebook, and the number of
entries in the codebook grows exponentially with the number of
antennas, simulation results are not shown for $N_t > 12$.  As
expected, both the upper and lower bounds grow at the same rate as
$N_t$ increases.

%--------------------------------------------
\subsection{Asymptotic Behavior}
%-------------------------------------------

We now study the behavior of the optimal $T, B$ and $D$,
and the capacity as $N_t \to \infty$.
With $D$ transmitted symbols in an $L$-symbol packet the effective
capacity $\mC = (\D/\Lb) C$ where $\D = D/N_t$ and $\Lb = L/N_t$.  The
associated bounds are $\mC_u = (\D/\Lb) C_u$ and $\mC_l = (\D/\Lb)
C_l$.  From Theorem~\ref{bnd} and \eqref{vw}, we can write $\mC_l$ and
$\mC_u$ as functions of $\{ \T, \B, \D \}$ and optimize, i.e.,
for the lower bound we wish to
\begin{align}
   \max_{\T,\B,\D} \ & \mC_l \\
   \text{subject to} \quad &\T + \mu \B + \D = \Lb . \label{tbd}
\end{align}
Let $\{ \T^o_l, \B^o_l, \D^o_l \}$ denote the optimal values of $\T$,
$\B$, and $\D$, respectively, and let $\mC_l^o$ denote the maximized
lower bound on capacity.  Similarly, maximizing the upper bound
gives the optimal parameters
$\{ \T^o_u, \B^o_u, \D^o_u \}$ and the corresponding bound $\mC_u^o$.
These optimized values can be easily computed numerically, and
also allow us to characterize the asymptotic
behavior of the {\em actual} capacity.\footnote{In what follows
all logarithms are assumed to be natural.}
\begin{theorem}
\label{limit}
Let $\{ \T^o, \B^o, \D^o \} = \arg \max_{ \{\T,\B,\D \} } \mC$
subject to \eqref{tbd}. As $N_t \to \infty$,
  \begin{align}
    & \T^o \log N_t \to \Lb \label{to}\\
    & \B^o \log N_t \to \frac{1}{\mu} \Lb \label{bo}
%    & \frac{\D^o} = \Lb \left( 1 - c(N_t) \right) \label{do}
  \end{align}
%where $c(N_t) \log N_t/2 \to 1$,
and the capacity satisfies
\begin{equation}
  \mC^o - \log(\rho N_t) + 2 \log \log N_t \to \zeta
 \label{rno}
\end{equation}
where $\zeta$ is a constant bounded by
\begin{equation}
  \zeta^* - \log(1 + \rho) \le \zeta \le \zeta^*
\end{equation}
where $\zeta^* = \log ( \Lb^2 \log(2)) - \log(\mu (1 + \rho^{-1})) - 2$.
\end{theorem}

The proof is given in Appendix~\ref{opt_cap_miso}.  Combining
\eqref{to} and \eqref{bo} with \eqref{tbd} gives the corresponding
behavior of the data segment
\begin{equation}
    \frac{\D^o}{\Lb} = 1 - \delta (N_t)
\label{do}
\end{equation}
where $\delta (N_t) \log N_t/2 \to 1$.

According to the theorem, as $N_t$ becomes large, to maximize the
achievable rate the fraction of $\Lb$ devoted to training and feedback
tends to zero, in which case the rate increases as 
$\log (\rho N_t) - 2\log \log N_t$.
The achievable rate with RVQ and perfect channel estimation is 
$E[\log (1+\rho \| \bh \|^2)]$, which grows as $\log(\rho N_t)$.  
Hence the loss of $2 \log \log N_t$ is due to imperfect 
channel estimation.\footnote{The capacity estimate in the theorem
becomes accurate when $N_t$ is large enough so that 
$\bar{L}/\log N_t$ is small, in which case
the loss term $2\log \log N_t$ is greater than
the constant offset $\zeta$.}
Theorem~\ref{limit} also implies that $\mu B /T \to 1$, i.e., the
fraction of the packet devoted to feedback is asymptotically the same
as that for training. This equal allocation therefore balances the
reductions in capacity due to estimation and quantization.

The preceding analysis applies if the beamforming
vectors during training are chosen to be unit vectors. Namely, the
matrix $\bVT$ can be taken to be diagonal, which corresponds to
transmitting the sequence of training symbols over the transmit
antennas successively one at a time.  Hence the fact that the optimal
$T$ increases as $N_t / \log N_t$ implies that only $N_t /\log N_t$
antennas are activated.  Since $\Lb = L/N_t$ is fixed, we
conclude that as the coherence time $L$ increases, the optimal number of
transmit antennas should increase as $L / \log L$.
The training and feedback overhead therefore reduces the
number of antennas that can be effectively used by a
factor of $1/\log L$.

%%%%%%%%%%%%%%%%%%%%%%%%%%%%%%%%%%%%%%%%%%%%%%%%%%%%%%%%%%%%%%%%%%%%%
\section{Multi-Input Multi-Output Channel}
\label{tmimo}
%%%%%%%%%%%%%%%%%%%%%%%%%%%%%%%%%%%%%%%%%%%%%%%%%%%%%%%%%%%%%%%%%%%%%%

In this section, we let the number of receive antennas $N_r$
scale with $N_t$.
As for the MISO channel, we can bound the capacity with limited
training and feedback as follows,
\begin{align}
  C & \le C_u = \log( 1 + \rho \varw + \rho E [\eta]) \label{CCu} \\
  C & \ge C_l = (1 - \cNt)\log \left(1 + \frac{\rho}{1 + \rho \varw}
    E [\eta] \right) \label{CCl}
\end{align}
where $\eta = \bvHh^{\dag} \bHh^{\dag} \bHh \bvHh$ and 
\begin{equation}
\cNt = \frac{ \sigma_{\eta} }{ 2  E [\eta] }
\label{cNt}
\end{equation}  
where $\sigma_{\eta}$ is the standard deviation of $\eta$.

We would like to express the bounds \eqref{CCu} and \eqref{CCl}
as functions of $\T$ and $\B$.  
%Reference \cite{hassibi} shows that
%the set of beamforming vectors, which achieves the Welch bound,
%minimizes the MSE.  
As discussed in Section \ref{sys_mod},
the variance of the estimation error is again given by \eqref{vw}.  
Although it is difficult to evaluate $E [\eta]$ explicitly for 
finite $(N_t, N_r, B)$, it can be computed in the large system limit
as the parameters tend to infinity with fixed ratios
$\N=N_r/N_t$ and $\B$. Specifically, since $\bHh$ has 
{\em i.i.d.} elements with variance $1 -\varw$, we have
\begin{equation}
  \frac{1}{N_t} \eta  \longrightarrow (1 - \varw) \srvq
\label{1Nte}
\end{equation}
in the mean square sense, where the asymptotic received signal
power with RVQ $\srvq$ is evaluated in \cite{SanHon09}, and is a 
function of $\N$ and $\B$.  Therefore
\begin{equation}
  E [ \eta ] = (1 - \varw) \srvq N_t + \kappa (N_t)
\label{Eta}
\end{equation}
where $\kappa (N_t) / N_t \to 0$. Characterizing
$\kappa ({N_t})$ explicitly appears to be difficult, but 
this is not needed to prove the following 
theorem.\footnote{We will assume that $\kappa (N_t)$
is a smooth function of $\T$ and $\B$ for all $N_t$, and 
that $\kappa(N_t)/N_t$ converges to zero uniformly over all $\T$ and $\B$.}
Substituting \eqref{Eta} and \eqref{vw} into \eqref{CCu} 
and \eqref{CCl} gives upper and lower bounds on the capacity, 
$C_l$ and $C_u$, respectively, as functions of $\T$ and $\B$.
Maximizing both bounds over $\T$ and $\B$
%\begin{gather}
%  \max_{\T, \B, \D} \ \mC_l = \frac{\D}{\Lb} C_l , \\
%  \max_{\T, \B, \D} \ \mC_u = \frac{\D}{\Lb} C_u , \\
%\text{subject to} \quad \T + \mu \B + \D = \Lb . \label{subm}
%\end{gather}
leads to the following theorem, which characterizes the 
asymptotic behavior of the actual capacity.
\begin{theorem}
\label{mimo_limit}
Let $\{ \T^o, \B^o, \D^o \} = \arg \max_{ \{\T,\B,\D \} } \mC$ subject
to \eqref{tbd}. As $(N_t,N_r) \to \infty$ with fixed $\N = N_r / N_t$,
\begin{align}
\label{tomimo}
 &\T^o \log N_t \longrightarrow \Lb \\ 
\label{bomimo}
 &\B^o \log^2 N_t  \longrightarrow  \frac{\Lb^2 \log 2}{2 \mu^2 \N}
% & \frac{\D^o}{1 - \frac{1}{\log N_t} - \left( \frac{\Lb \log 2}{2 \N
%  \mu} \right) \frac{1}{\log^2 N_t}} \longrightarrow  \Lb
\end{align}
and the capacity satisfies
\begin{equation}
  \mC^o - \log(\rho N_t) + \log \log N_t \to \xi
\end{equation}
where 
\begin{equation}
  \xi^* - \log(1 + \rho) \le \xi \le \xi^*
\end{equation}
and $\xi^* = \log ( \Lb \N) - \log(1+\rho^{-1}) - 1$.
\end{theorem}

The proof is given in Appendix~\ref{opt_cap_mimo}.
Combining \eqref{tomimo}, \eqref{bomimo}, and \eqref{tbd}
gives the corresponding behavior of the optimized data segment
\begin{equation}
\frac{\D^o}{\Lb} = 1 - \epsilon_1 (N_t) - \epsilon_2 (N_t)
\label{domimo}
\end{equation}
where $\epsilon_1 (N_t) \log N_t \to 1$
and $\frac{2 \N \mu}{\Lb \log 2} \epsilon_2 (N_t) \log^2 N_t \to 1$.

Theorem \ref{mimo_limit} states that the optimal
training length for the MIMO channel grows as
$N_t / \log N_t$, which is the same as for the MISO channel.  
Hence as $N_t$ becomes large, 
only $N_t / \log N_t$ transmit antennas should be activated.
(All receive antennas are used, since this does
not change the training overhead.)

Theorem \ref{mimo_limit} also states that the capacity with limited
training and feedback increases as $\log(\rho N_t) - \log \log N_t$.
For large $N_t$ the loss in achievable rate due to training and
feedback therefore increases as $\log \log N_t$, as opposed to $2\log
\log N_t$ for the MISO channel.  This gain is due to the smaller MIMO
feedback overhead.  Namely, because of the additional antennas for the
MIMO channel, the optimal normalized feedback length tends to zero at
the rate $1 / \log^2 N_t$, as opposed to $1/\log N_t$ for the MISO
channel.  Note, however, that the training overhead is the same since
the same training symbols are used to estimate the channel gains to
all receive antennas simultaneously.  Hence the ratio of optimized
feedback to training overhead for the MIMO channel $\frac{\mu
  \B^o}{\T^o} \to 0$ as $1/\log N_t$.

%%%%%%%%%%%%%%%%%%%%%%%%%%%%%%%%%%%%%%%%%%%%%%%%%%%%%%%%%%%%%%%%%%%%%%%%
\section{Numerical Results}
\label{numerical}
%%%%%%%%%%%%%%%%%%%%%%%%%%%%%%%%%%%%%%%%%%%%%%%%%%%%%%%%%%%%%%%%%%%%%%%%

\begin{figure}
  \centering
  \includegraphics[width=3.25in]{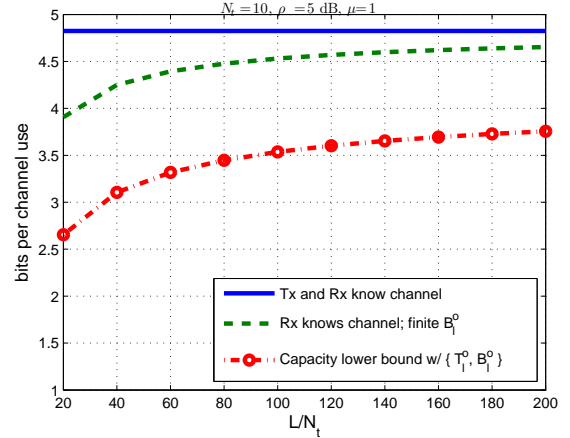}
  \caption{Achievable rate versus normalized packet length $\Lb$.}
  \label{capVSl}
\end{figure}

Fig.~\ref{capVSl} shows achievable rates for the MISO channel versus
normalized coherence time $\Lb= L/N_t$ with
different assumptions about channel knowledge 
at the transmitter and receiver.
Three curves are shown: (1) the optimized lower bound
on capacity $\mathcal{C}_l^o$, (2) the capacity assuming the
receiver knows the channel, but with a quantized beamformer,
and (3) the capacity with perfect channel knowledge at
the transmitter and recevier (optimal beamforming).
Parameters are $N_t=10$, $\rho= 5$ dB, and $\mu=1$ (BPSK feedback).
As expected, the gaps between the curves diminishes to zero
with increasing coherence time, albeit slowly.
This reflects the fact that the training and feedback
overhead tends to zero as $1/\log L$.

\begin{figure}
  \centering \includegraphics[width=3.25in]{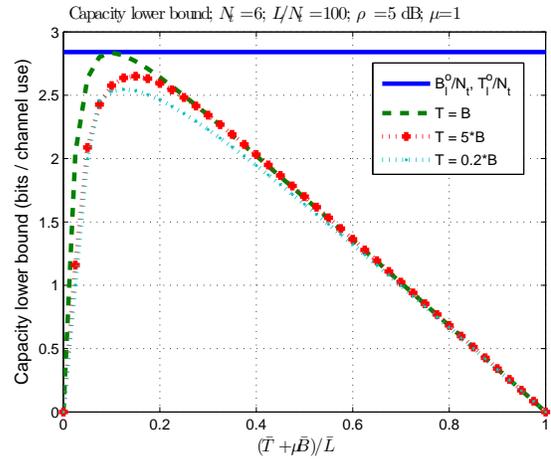}
  \caption{Lower bound on capacity versus normalized training
  and feedback $(\T + \mu \B)/\Lb$ with different allocations $\T/(\mu \B)$.}
  \label{capVStb}
\end{figure}

Fig.~\ref{capVStb} illustrates the sensitivity of the capacity
for the MISO channel to different choices for training
and feedback overhead.
The lower bound $\mathcal{C}_l^o$ is plotted
versus the fractional overhead $(\T + \mu \B)/\Lb$
with different relative allocations $\T/(\mu \B)$.
Parameters are
$\Lb = 100$, $N_t = 6$, $\mu = 1$, and $\rho = 5 \ \text{dB}$.
The solid line corresponds to optimized overhead $T_l^o$ and $B_l^o$.
The capacity is zero when
$\T + \B =0$, since the estimate is uncorrelated
with the channel, and when $\T + \B = \Lb$, since $\D=0$.
With equal amounts of training and feedback the rate
is essentially equal to that with optimized parameters.
The peak is achieved when $(\T + \B)/\Lb = 0.1$.
The performance is relatively robust to this choice, i.e.,
small deviations from this value result in a relatively
small performance loss, although the performance loss
increases substantially as the deviations become larger.
Likewise, the figure also shows that there is a
significant performance degradation when $\B$ deviates
significantly from $\T$.

\begin{figure}
  \centering
  \includegraphics[width=3.25in]{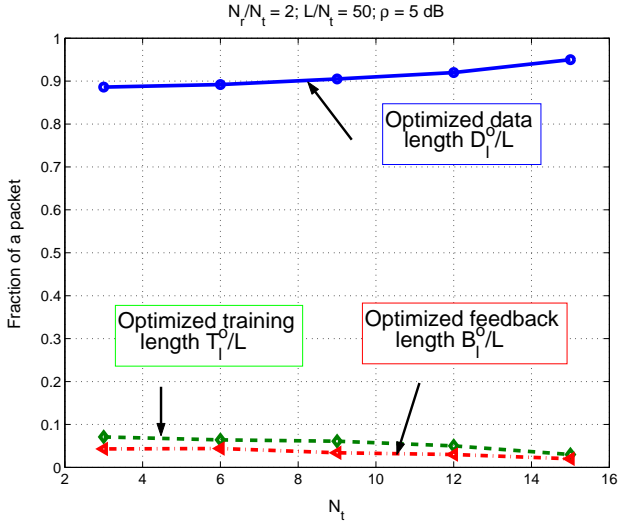}
  \caption{Optimized training and feedback overhead, and
fraction of data symbols $\{\T^o_l/\Lb,\B^o_l/\Lb,\D^o_l/\Lb \}$
versus number of transmit antennas $N_t$.}
  \label{opt_sol_mimo}
\end{figure}

\begin{figure}
  \centering
  \includegraphics[width=3.25in]{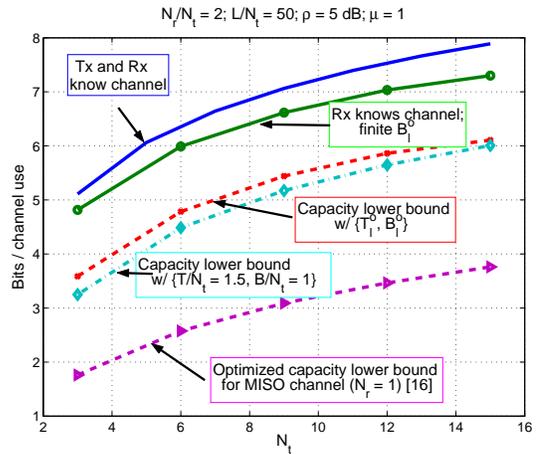}
  \caption{Achievable rate for MIMO channel versus number of transmit
  antennas $N_t$ with different assumptions about channel knowledge at
  the receiver and transmitter.  Also shown is the optimized capacity
  lower bound for the corresponding MISO channel}
  \label{rateVSnt_mimo}
\end{figure}

The optimized training, feedback, and data portions of the packet
(normalized by the packet length $L$) versus $N_t$ for the MIMO channel
are shown in Fig.~\ref{opt_sol_mimo}. These values were obtained by
numerically optimizing the capacity lower bound, and are therefore
denoted as $B^o_l$, $T^o_l$, and $D^o_l$ in the figure.  System
parameters are $\N = 2$, $\Lb = 50$, $\mu = 1$, and $\rho = 5$ dB.  As
predicted by Theorem~\ref{mimo_limit}, both the optimal $\T$ and $\B$
decrease to zero, with $\B$ decreasing somewhat faster than $\T$.
The associated capacity lower bound is shown in Fig.~\ref{rateVSnt_mimo}.
Also shown is the
capacity lower bound with the heuristic choice of parameters $\B = 1$
(one feedback bit per coefficient)
and $\T = 1.5$ (1.5 training symbols per coefficient).
For $N_t = 3$, the bound with
optimized parameters is approximately 10\% greater than that with the
heuristic choice.  Those results are compared with the capacity with
perfect channel knowledge at {\em both} the transmitter and receiver,
and the capacity with perfect channel knowledge at the receiver {\em
only} with $B^o_l$ feedback bits. This comparison
indicates how much of the loss in achievable rate for the model
considered is due to channel estimation at the receiver
(including associated overhead), and how much is due
to quantization of the precoding matrix.

The results show that for $N_t = 3$, the capacity with perfect channel
knowledge at both the transmitter and receiver is about 40\% larger
than the rate with optimized feedback and training lengths.  Knowing
the channel at the receiver achieves most of this gain, largely due to
the elimination of associated training overhead.  Of course, this gap
tends to zero as the block size $\bar{L}\to\infty$.  Also shown in the
figure for comparison is the capacity lower bound for a MISO channel
with optimized training and feedback lengths.  This is substantially
lower than that shown for the MIMO channel.  From
Theorems~\ref{limit} and~\ref{mimo_limit} the gap between
the optimized lower bounds for the MISO and MIMO channels increases as
$\log \log N_t$.

\begin{figure}
  \centering \includegraphics[width=3.25in]{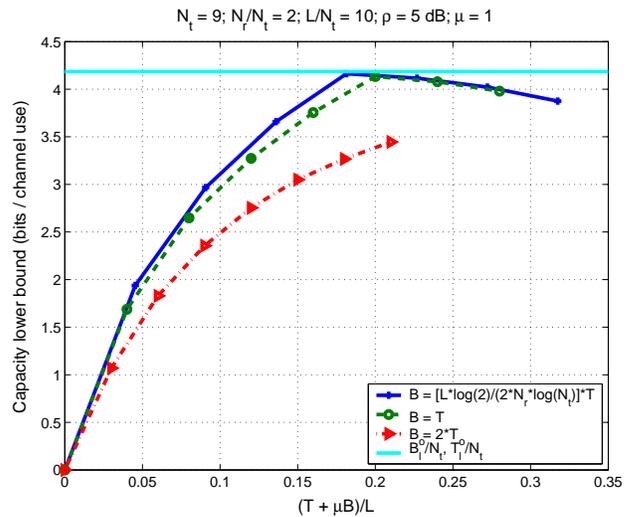}
  \caption{Lower bound on beamforming capacity for MIMO channel
versus normalized training and feedback $(\T + \mu \B)/\Lb$.}
  \label{capVStb_mimo}
\end{figure}

Similar to Fig.~\ref{capVStb},
Fig.~\ref{capVStb_mimo} shows the capacity lower bound
versus total overhead $(\T + \mu \B)/\Lb$ for a MIMO channel.
The solid line corresponds to optimized parameters
with $\Lb = 10$, $N_t = 9$, $\N = 2$, $\mu = 1$, and $\rho = 5$ dB.
The curves are obtained by numerical optimization.  For the case
considered, these results show that the rate achieved with equal
portions of training and feedback is close to the maximum
(corresponding to optimized training and feedback).  Allocating the
overhead according to the asymptotic results in
Theorem~\ref{mimo_limit}, i.e., taking $\mu \B / \T = \Lb \log 2/(2
\mu \N \log N_t)$, performs marginally better than allocating equal
training and feedback.  The total optimized overhead in this case is
$(\T + \B)/\Lb \approx 0.2$.  The performance degrades when $\B$
deviates significantly from $\T$ (as shown by the curve
corresponding to $\B = 2 \T$).  (The three
curves shown are not extended to $(\T +\B)/ \Lb = 1$ since the
simulation complexity associated with RVQ increases
exponentially with $\B$.) Compared with the results for
the MISO channel in Fig.~\ref{capVStb}, the capacity
for the MIMO channel is somewhat more robust with respect
to variations in overhead.

%%%%%%%%%%%%%%%%%%%%%%%%%%%%%%%%%%%%%%%%%%%%%%%%%%%%%%%%%%%%%%%%%%%
\section{Conclusions}
\label{conclude}
%%%%%%%%%%%%%%%%%%%%%%%%%%%%%%%%%%%%%%%%%%%%%%%%%%%%%%%%%%%%%%%%%%%

We have presented bounds on the capacity of both MISO and MIMO block
Rayleigh fading channels with beamforming, assuming limited training
and feedback.  For a large number of transmit antennas, we have
characterized the optimal amount of training and feedback as a
fraction of the packet duration, assuming linear MMSE estimation of
the channel, and an RVQ codebook for quantizing the beamforming
vector.  Our results show that the optimized training length for both
MISO and MIMO channels increases as $N_t/ \log N_t$, which can be
interpreted as the optimal number of transmit antennas to
activate. The ratio of optimized feedback to training overhead tends
to one for the MISO channel, but tends to zero as $1/\log N_t$ for the
MIMO channel, since additional receiver antennas improve robustness
with respect to quantization error.  The loss in capacity due to
overhead increases as $\log \log N_t$ for the MIMO channel, and as $2
\log \log N_t$ for the MISO channel.  

Although the pilot scheme considered is practical, it is most
likely suboptimal.  That is, in the absence of feedback such a
pilot-based scheme is strictly suboptimal, although it is nearly
optimal at high SNRs \cite{hassibi}.  Computing the capacity of
the block fading channel considered with feedback
and no channel knowledge at the receiver and transmitter
is an open problem. Consequently, although the optimal (capacity-maximizing)
number of transmit antennas should still be limited by the coherence time,
the growth rate may differ from the $L/\log L$ growth rate
shown here for the pilot scheme.

The model and analysis presented here can be extended in a few different
directions. A natural generalization of the MIMO beamforming model 
is to allow a general transmit precoding matrix with
rank greater than one. The additional overhead should impose
a limit on both the number of beams and antennas that can
effectively be used. Also, the powers allocated to the training
and data portions of the coherence block can be optimized
in addition to the fraction of overhead symbols.
Finally, feedback and training overhead becomes especially
important in multi-user MIMO scenarios, such as the cellular downlink.
The optimal overhead scaling with coherence time in those scenarios
remains to be studied.

%*********************************************************************
\appendix
%*********************************************************************

%%%%%%%%%%%%%%%%%%%%%%%%%%%%%%%%%%%%%%%
\subsection{Proof of Lemma~\ref{enu}}
\label{snr_bnds}
%%%%%%%%%%%%%%%%%%%%%%%%%%%%%%%%%%%%%%

We need to evaluate \eqref{mean}.  Letting $n = 2^B$, we first bound
\begin{align}
&n B \left(n, 1 + \frac{1}{N_t - 1} \right) \nonumber\\
   & = \frac{n \Gamma(n) \Gamma \left(1 + \frac{1}{N_t - 1} 
     \right)}{ \Gamma \left( n + 1 + \frac{1}{N_t - 1} \right)}\\
   & = \Gamma \left( 1 + \frac{1}{N_t - 1} \right) \frac{\Gamma (n + 2)}
     {(n + 1) \Gamma \left(n +1 +\frac{1}{N_t-1} \right)}\\
   & \ge \Gamma \left( 1 + \frac{1}{N_t - 1} \right) 
   (n + 1)^{- \frac{1}{N_t - 1}}  \label {bnb}\\
   & = \Gamma \left( 1 + \frac{1}{N_t - 1} \right) \left( 1 +
   \frac{1}{n} \right)^{-
   \frac{1}{N_t - 1}} 2^{-\B \left(1 + \frac{1}{N_t - 1} \right)}
   \label{f1nt}
\end{align}
where we have used $B (p,q) = \Gamma(p) \Gamma(q) /
\Gamma(p+q)$, the identity $\Gamma( k + 1 ) = k \Gamma(k)$ for $k
\in \mathbb{N}$, and the inequality $\Gamma(k + 1)/\Gamma(k + x )
\ge k^{1 - x}$ for $0 \le x \le 1$ \cite{kershaw}.  
Since $\Gamma(x)$ is convex for $x \in [1,2]$, for $N_t
\ge 2$,
\begin{equation}
  \Gamma \left( 1 + \frac{1}{N_t - 1} \right) \ge \Gamma(1) +
  \frac{\Gamma'(1)}{N_t-1} = 1 - \frac{\gamma}{N_t-1}
\label{Gl1}
\end{equation}
where $\gamma = 0.5772\ldots$ is the Euler constant.  Expanding the
second factor on the right-hand side of \eqref{f1nt} in a Taylor
series gives
\begin{align} 
  \left( 1 + \frac{1}{n} \right)^{- \frac{1}{N_t - 1}}
  & = 1 - \frac{1}{N_t-1} \frac{1}{n} + \frac{N_t}{2! (N_t-1)^2}
  \frac{1}{n^2} \nonumber \\
  & \quad - \frac{N_t(2 N_t - 1)}{3! (N_t-1)^3} \frac{1}{n^3} +
   \cdots \label{1f1}\\
  & \ge 1 - \frac{1}{n(N_t-1)}  \label{ge1}
\end{align}
since the magnitude of each term in \eqref{1f1} is decreasing.  We
also expand
\begin{align}
  &(2^{-\B})^{\frac{1}{N_t-1}} \nonumber \\
    & = 1 - \frac{1}{N_t-1}(1 - 2^{-\B}) - \frac{N_t -2}{2! (N_t-1)^2}
    (1 - 2^{-\B})^2 \nonumber \\
    & \quad - \frac{(N_t -2)(2 N_t - 3)}{3! (N_t-1)^3}
    (1 - 2^{-\B})^3 - \cdots \\
    & \ge 1 - \frac{1}{N_t-1} \big[ (1 - 2^{-\B}) + (1 - 2^{-\B})^2 \nonumber \\
    & \quad + (1 - 2^{-\B})^3 + \cdots \big] \\
    & = 1 - \frac{1}{N_t-1} (2^{\B} - 1) .  \label{1frc}
\end{align}
Substituting \eqref{Gl1}, \eqref{ge1}, and \eqref{1frc} into
\eqref{f1nt} yields
\begin{align}
  & n B  \left(n, 1 + \frac{1}{N_t - 1} \right) \nonumber \\
    & \ge 2^{-\B} \left( 1 - \frac{\gamma}{N_t-1} \right) \left(
    1 - \frac{1}{n(N_t-1)} \right) 
    \left( 1 - \frac{2^{\B} - 1}{N_t-1}\right)\\
    & \ge 2^{-\B} \left[ 1 - \frac{1}{N_t-1} ( 2^{\B} 
      - 1 + \gamma + 2^{-B}) \right] . \label{ge2B}
\end{align}
The inequality \eqref{ge2B} holds for $N_t \ge 2$ and $\B \ge 0$.
Therefore
\begin{align}
\label{eq:Enu}
E [ \nu ] & = 1 -  2^B B  \left(2^B, 1 + \frac{1}{N_t - 1} \right) \\
          & \le 1 -  2^{-\B} + \frac{1 + (\gamma - 1) 2^{-\B} +
	  2^{-\B N_t}}{N_t -1} .
\end{align}

To show \eqref{12b2}, we derive the following upper bound
\begin{align}
   & n B \left(n, 1 + \frac{1}{N_t - 1} \right) \nonumber \\
  & = \Gamma \left( 1 + \frac{1}{N_t - 1} \right) \frac{\Gamma (n + 1)}
     {\Gamma \left(n +1 +\frac{1}{N_t-1} \right)} \\
  & \le \Gamma \left( 1 + \frac{1}{N_t - 1} \right) \left( n +
  \frac{N_t}{2 (N_t - 1)} \right)^{-\frac{1}{N_t - 1}} \label{leG}\\
  & = \Gamma \left( 1 + \frac{1}{N_t - 1} \right) \left( 1 +
  \frac{N_t}{2 n (N_t - 1)} \right)^{-\frac{1}{N_t - 1}}
  2^{-\frac{\B}{N_t -1}} 2^{-\B} . \label{glf}
\end{align}
The inequality \eqref{leG} is shown in \cite{wolfram}.  Since every
factor in \eqref{glf} is less than or equal to one, we conclude that
\begin{equation}
  n B  \left(n, 1 + \frac{1}{N_t - 1} \right) \le 2^{-\B} ,
\end{equation}
and combining with \eqref{eq:Enu} gives the lower bound \eqref{12b2}.

%%%%%%%%%%%%%%%%%%%%%%%%%%%%%%%%%%%%%%%%%%
\subsection{Proof of Lemma~\ref{l_bental}}
\label{pf_bental}
%%%%%%%%%%%%%%%%%%%%%%%%%%%%%%%%%%%%%%%%%%

Since $\log \left(1 + \frac{1}{\varz} X \right)$ is concave for $X \in
[0, \infty)$ and
\begin{equation}
\lim_{t \to \infty} \frac{1}{t} \log \left(1 + \frac{1}{\varz}t \right)
= 0 ,
\end{equation}
we can apply the following inequality in \cite{bental}
\begin{multline}
E \left[ \log \left(1 + \frac{1}{\varz} X \right) \right] \\
  \ge
  \left(1 - \frac{E \left| X - E [ X ] \right|}{2 E [ X
  ]}\right) \log \left(1 + \frac{1}{\varz} E[ X ] \right) .
\end{multline}
Now set $X = A \nu$, where $A \triangleq \| \bhh \|^2$ and 
$\nu \triangleq | \bhh \bvhh |^2 / \| \bhh \|^2$.
Since $A$ and $\nu$ are independent, and using
the relation
$\left( E \left| X - E [ X ] \right| \right)^2 \le  \var [ X ]$,
we obtain
\begin{align}
  \frac{E \left| X - E [ X ] \right|}{2 E [ X ]} 
    & \le \frac{\sqrt{ \var [ X ]}}{2 E [ X ]} \\
    & = \frac{1}{2} \sqrt{ \frac{E [ A^2 ]}{E^2 [ A ]} 
      \frac{E [ \nu^2 ]}{ E^2 [ \nu ]} -1 } .
\label{EvlX}
\end{align}

Each element in $\bhh$ is {\em i.i.d.} with a complex Gaussian distribution.
Hence $A$ is Gamma distributed so that
\begin{equation}
  \frac{E [ A^2 ]}{E^2 [ A ]} = 1 + \frac{1}{N_t} .
\label{sEa}
\end{equation}
To evaluate $E [ \nu^2] / E^2 [ \nu ]$ in \eqref{EvlX}
we first compute
\begin{align}
 & E [ (1 - \nu)^2 ] \nonumber \\
   & = \int_0^1 ( 1 - v )^2 f_{\nu} (v)\, \diff v \\
   & = \int_0^1 ( 1 - v )^2 \big[ n (N_t-1) \left( 1 - (1 -
   v)^{N_t-1} \right)^{n-1} \nonumber\\
   & \quad \times (1 - v)^{N_t - 2} \big]\, \diff v
\end{align}
where $f_{\nu} ( \cdot )$ is the pdf for $\nu$, and is
given in \cite{chun_love}.  Applying the change of variables
$q = (1 - v)^{N_t-1}$ gives
\begin{align}
E [ (1 - \nu)^2 ] & = n \int_0^1 q^{\frac{2}{N_t-1}} (1 - q)^{n-1} \,
                  \diff q \\
              & = n B \left( n, 1 + \frac{2}{N_t - 1} \right) .
\end{align}
Therefore
\begin{align}
\var [ \nu ] & = E[ \nu^2 ] - E^2 [ \nu] \\
             & = n B \left( n, 1 + \frac{2}{N_t - 1} \right) 
             - (1 - E [\nu])^2\\
             & = n B \left( n, 1 + \frac{2}{N_t - 1} \right) - n^2 B^2
             \left( n, 1 + \frac{1}{N_t - 1} \right) . \label{nBl}
\end{align}
Applying the inequality in \cite{wolfram}, we have
\begin{align}
& n B \left( n, 1 + \frac{2}{N_t - 1} \right) \nonumber \\
  & = \Gamma \left( 1 + \frac{2}{N_t - 1} \right) \frac{\Gamma (n + 1)}
     {\Gamma \left(n +1 +\frac{2}{N_t-1} \right)} \\
  & \le  \Gamma \left( 1 + \frac{2}{N_t - 1} \right) \left( n +
  \frac{1}{N_t-1} + \frac{1}{2} \right)^{-\frac{2}{N_t-1}} .
\label{lGl}
\end{align}
Substituting \eqref{lGl} and \eqref{bnb} into \eqref{nBl} gives
\begin{align}
& \var[ \nu ] \nonumber \\
& \le \Gamma \left( 1 + \frac{2}{N_t - 1} \right) \left( n +
  \frac{1}{N_t-1} + \frac{1}{2} \right)^{-\frac{2}{N_t-1}} \nonumber \\
& \quad - 
  \Gamma^2 \left( 1 + \frac{1}{N_t - 1} \right) \left( n + 1
  \right)^{-\frac{2}{N_t-1}} \\
& = 2^{-2 \B \left(1 + \frac{1}{N_t-1} \right)} \bigg[ \Gamma 
  \left( 1 + \frac{2}{N_t - 1} \right) \nonumber \\
& \quad \times \left( 1 +
  \frac{1}{n(N_t-1)} + \frac{1}{2n} \right)^{-\frac{2}{N_t-1}}
  \nonumber \\
& \quad - 
  \Gamma^2 \left( 1 + \frac{1}{N_t - 1} \right) \left( 1 + \frac{1}{n}
  \right)^{-\frac{2}{N_t-1}} \bigg]\\
& \le 2^{-2 \B \left(1 + \frac{1}{N_t-1} \right)} \bigg[ \Gamma 
  \left( 1 + \frac{2}{N_t - 1} \right) \nonumber \\
& \quad - \Gamma^2 
  \left( 1 + \frac{1}{N_t - 1} \right) \left( 1 + \frac{1}{n}
  \right)^{-\frac{2}{N_t-1}}\bigg] .
\label{l22}
\end{align}
Since the second factor in \eqref{glf} is less than or equal to one,
we have
\begin{equation}
E [ \nu ] \ge 1 - \Gamma \left( 1 + \frac{1}{N_t - 1} \right) 
  2^{-\B \left(1 + \frac{1}{N_t-1} \right)}.
\label{Enug}
\end{equation}
Finally, combining \eqref{EvlX}, \eqref{sEa}, \eqref{l22}, and
\eqref{Enug} gives $E \left| X - E [ X ] \right|/(2 E [ X ]) \le \dNt
$ in \eqref{dNt}, which completes the proof.

%%%%%%%%%%%%%%%%%%%%%%%%%%%%%%%%%%%%%%%%%%
\subsection{Proof of Theorem~\ref{limit}}
\label{opt_cap_miso}
%%%%%%%%%%%%%%%%%%%%%%%%%%%%%%%%%%%%%%%%%

We first maximize the upper bound given by
\begin{align}
  \mC_u & =  \frac{\D}{\Lb} C_u \\
     & = \frac{\D}{\Lb} \log \left(  \frac{\rho}{1 + \rho^{-1}} \T ( 1
     - 2^{-\B} ) N_t\right) + \frac{\D}{\Lb} \log (1 + \rNt) \label{Cfra}
\end{align}
where
\begin{equation}
  \rNt = \frac{(1 + \rho^{-1})^2 - \T}{\T (1 - 2^{-\B}) N_t} + \frac{1
  + (\gamma-1)2^{-\B} + 2^{-\B N_t}}{(N_t-1)(1 - 2^{-\B})} .
\label{rnt}
\end{equation}
The expression for $\sigma_w^2$ in \eqref{vw}
with $\T \leq 1$ has been used in \eqref{Cfra}, 
since we will show that $\T \to 0$ as $N_t \to \infty$.
We wish to characterize the behavior of the optimal parameters
$\{ \Tu, \Bu, \Du \}$ as $N_t \to \infty$.

The Lagrangian is given by
\begin{equation}
  \mL = \mC_u + \lambda (\Lb - \T -\mu \B -\D)
\end{equation}
where $\lambda$ is the Lagrangian multiplier.  
Setting the partial derivatives of $\mL$
with respect to $\D$, $\T$, $\B$, and $\lambda$ to zero
gives the necessary conditions
\begin{multline}
\label{p1} 
  \log \left( \frac{\rho}{1 + \rho^{-1}} \right) + \log ( \T ) + \log
     ( 1 - 2^{-\B}) + \log N_t \\ + \log( 1 + \rNt) 
     - \Lb \lambda = 0
\end{multline}
\begin{gather}
  \frac{\D}{\T} + \left( \frac{\D}{1 + \rNt} \right)
     \frac{\partial \rNt}{\partial \T} - \Lb
     \lambda = 0 \label{p2}\\
  \frac{\D \log 2}{2^{\B} - 1} + \left( \frac{\D}{1 + \rNt} \right)
     \frac{\partial \rNt}{\partial \B} - \Lb \mu
     \lambda = 0 \label{p3}\\
  \Lb - \T - \mu \B - \D = 0  . \label{p4}
\end{gather}
Substituting \eqref{p2}, \eqref{p4}, and the expression for
$\frac{\partial \rNt}{\partial \T}$ into \eqref{p1} gives
\begin{multline}
\label{tlgn}
  \T \log N_t + \T \log \left( \frac{\rho}{1 + \rho^{-1}} \right) +
  \T \log ( 1 - 2^{-\B}) \\ + \T \log \T + \T \log( 1 + \rNt) \\
  = (\Lb - \T - \mu \B) \left( 1 - \frac{(1+\rho^{-1})^2}{(1 + \rNt)(1 - 2^{-\B}) \T N_t}\right) .
\end{multline}
We first observe that $(1 - 2^{-\Bu}) \Tu N_t \to \infty$
as $N_t \to \infty$.
Otherwise, it easily verified from \eqref{Cfra} that $C_u$ 
must be bounded by a constant. However, this is clearly
suboptimal, since if $\B$ and $\T$
are constants, then $C_u$ grows as $O(\log N_t)$.
This observation implies that $\rNt \to 0$.

As $N_t \to \infty$, the right-hand side of \eqref{tlgn} converges to
$\Lb - \T - \mu \B$, so that \eqref{tlgn} implies $\T \to 0$.
As $N_t \to \infty$, \eqref{tlgn} therefore implies
\begin{equation}
  \T \log N_t \to \Lb - \mu \B.
\label{TNt}
\end{equation}

Combining \eqref{p2} and \eqref{p3} gives
\begin{equation}
  \B = \frac{1}{\log 2} \log\left(1 + \frac{\log 2}{\mu} \T \left(
   \frac{1}{1 + \xi({N_t})} \right)\right)
\label{Bf12}
\end{equation}
where
\begin{equation}
  \xi({N_t}) = \frac{\T}{1+ \rNt}
  \left(\frac{\partial \rNt}{\partial \T} - \frac{1}{\mu}
  \frac{\partial \rNt}{\partial \B} \right).
\label{xi}
\end{equation}
Since $\T \to 0$, and $\rNt \to 0$ uniformly over $\T$ and $\B$ (so
that the derivatives in \eqref{xi} must also tend to zero), it follows
that $\xi(N_t ) \to 0$.  Hence for large $N_t$ \eqref{Bf12} implies
that
\begin{equation}
\B = \frac{1}{\mu} \T + O(\T^2) ,
\label{BToT}
\end{equation}
where we have used the Taylor expansion
$\log(1 + x) = x + O(x^2)$ for small $x$. 
Combining \eqref{TNt} and \eqref{BToT}, it follows that
\begin{gather}
  \Tu \log N_t \to \Lb , \label{Tul}\\ 
  \Bu \log N_t \to \frac{1}{\mu} \Lb . \label{Bul}
\end{gather}

Substituting the optimal parameters in the capacity upper bound
\eqref{Cfra} gives
\begin{multline}
\mC_u^o - \frac{\Du}{\Lb} \log(\rho N_t) - \frac{\Du}{\Lb} \log \Tu - 
  \frac{\Du}{\Lb} \log(1 - 2^{-\Bu}) \\
  =  - \frac{\Du}{\Lb} \log(1 + \rho^{-1}) +
\frac{\Du}{\Lb} \log(1 + \rNt)
\end{multline}
where $\mC_u^o$ denotes the optimal $\mC_u$.  Taking $N_t \to \infty$
gives
\begin{multline}
\mC_u^o - \log(\rho N_t) + 2 \log \log N_t \\ \to 
  \log(\Lb^2\log 2) - 2 - \log[\mu(1 + \rho^{-1})] .
\label{Cuo}
\end{multline}

Following similar steps to optimize the lower bound \eqref{eq:Cl} gives
\begin{gather}
   \T_l^o \log N_t \to \Lb , \\ 
   \B^o_l \log N_t \to \frac{1}{\mu} \Lb .
\end{gather}
(Here we must show that $\dNt$ in \eqref{dNt} tends
to zero uniformly over all $\T$ and $\B$.)
The optimized lower bound satisfies
\begin{multline}
\mC_l^o - \log(\rho N_t) + 2 \log \log N_t \\ 
  \to \log(\Lb^2\log 2) -
2 - \log[\mu(1 + \rho^{-1})] - \log(1+\rho).
\end{multline}

Since the optimized bounds grow with $N_t$ at the same rate, the
capacity must also grow at that rate.  Hence we conclude that the
parameters that maximize the capacity exhibit the asymptotic behavior
stated in the theorem.

%%%%%%%%%%%%%%%%%%%%%%%%%%%%%%%%%%%%%%%%%%%%%%
\subsection{Proof of Theorem~\ref{mimo_limit}}
\label{opt_cap_mimo}
%%%%%%%%%%%%%%%%%%%%%%%%%%%%%%%%%%%%%%%%%%%%%%

Similar to the proof of Theorem~\ref{limit} in
Appendix~\ref{opt_cap_miso}, we first optimize 
the upper bound given by
\begin{equation}
  \mC_u = \frac{\D}{\Lb} \log \left(  \frac{\rho}{1 + \rho^{-1}} \T
  \srvq N_t \right) + \frac{\D}{\Lb} \log (1 + \sNt)
\label{mCu}
\end{equation}
where
\begin{equation}
  \sNt = \frac{(1 + \rho^{-1})^2 + (1 + \rho^{-1}) \kappa (N_t) -
  \T}{\T \srvq N_t} ,
\end{equation}
and we have substituted $\varw = 1 - \T / (1 + \rho^{-1})$,
corresponding to $\T \le 1$, since
we will show that the optimal normalized training length 
$\Tu \to 0$ as $N_t \to \infty$.

The Lagrangian for this optimization problem is given by
\begin{equation}
  \mL = \mC_u + \lambda (\Lb - \T -\mu \B -\D)
\end{equation}
where $\lambda$ is the Lagrange multiplier. 
The first-order necessary conditions are
\begin{multline}
  \log \left( \frac{\rho}{1 + \rho^{-1}} \right) + \log ( \T ) + \log
     ( \srvq ) + \log N_t \\ + \log( 1 + \sNt) 
     - \Lb \lambda = 0 \label{m1}
\end{multline}
\begin{gather}
  \frac{\D}{\T} + \left( \frac{\D}{1 + \sNt} \right)
     \frac{\partial \sNt}{\partial \T} - \Lb
     \lambda = 0 \label{m2}\\
  \left( \frac{\D}{\srvq} \right)
     \frac{\partial \srvq}{\partial \B} 
     + \left( \frac{\D}{1 + \sNt} \right)
     \frac{\partial \sNt}{\partial \B} - \Lb \mu
     \lambda = 0 \label{m3}\\
  \Lb - \T - \mu \B - \D = 0  .\label{m4}
\end{gather}

Substituting \eqref{m2} and \eqref{m4} into \eqref{m1} gives
\begin{multline}
  \T \log N_t + \T \log \left( \frac{\rho}{1 + \rho^{-1}} \right) +
  \T \log ( \srvq ) \\+ \T \log(\T)  + \T \log( 1 + \sNt)
  \\= (\Lb - \T - \mu \B) \left( 1 + \left( \frac{\T}{1 + \sNt} \right)
     \frac{\partial \sNt}{\partial \T} \right).
\end{multline}
Using an argument analogous to that used to show that $(1 - 2^{-\Bu})
\Tu N_t \to \infty$ as $N_t \to \infty$ in
Appendix~\ref{opt_cap_miso}, we can show that as $N_t \to \infty$, $\T
\srvq N_t \to \infty$, which implies that $\sNt \to 0$
uniformly in $\T$ and $\B$, so that $\left( \frac{\T}{1 + \sNt}
\right) \frac{\partial \sNt}{\partial \T} \to 0$.  Taking $N_t \to
\infty$ therefore gives
\begin{equation}
  \Tu \log N_t - \Lb \to 0,
\label{Tulo}
\end{equation}
assuming that $\Bu \to 0$, which will be proved next.

Substituting \eqref{m2} into \eqref{m3} to eliminate $\lambda$ and
rearranging gives
\begin{multline}
\label{nc3}
  \srvq \left( \frac{\partial \srvq}{\partial \B} \right)^{-1} \\=
  \frac{\T}{\mu} \left[ 1 + \frac{\T}{1 + \sNt} \left( \frac{\partial
  \sNt}{\partial \T} - \frac{1}{\mu} \frac{\partial
  \sNt}{\partial \B} \right)\right]^{-1} .
\end{multline}
Since $\T \to 0$ and $\sNt \to 0$,
\begin{equation} 
\srvq \left( \frac{\partial \srvq}{\partial \B} \right)^{-1} \longrightarrow 0 .
\label{fpsp}
 \end{equation}

For $0 \le \B \le \B^*$ it is shown in~\cite[Theorem 3]{SanHon09}
that $\srvq$ satisfies (after some rearrangement)
\begin{equation}
  \left( - \frac{\srvq}{\N} \right) \me^{ - \srvq / \N } = -
  \frac{1}{\me} 2^{-\B / \N}
\label{Nrp}
\end{equation}
where $\B^*$ is given by
\begin{equation}
\B^* = \frac{1}{\log 2}\left( \N\log(\sqrt{\N}) - \N \log(1 +
  \sqrt{\N}) + \sqrt{\N} \right).
\label{bst}
\end{equation}
We can therefore write $-\srvq /\N = W( - \frac{1}{\me} 2^{-\B /
  \N})$, where $W(x)$ is the Lambert-$W$ function.  It is
straightforward to show that
\begin{equation}
\srvq \left( \frac{\partial \srvq}{\partial \B} \right)^{-1} 
= \left( \frac{\partial [ \log \srvq]}{\partial \B} \right)^{-1} =
\frac{\srvq - \N}{\log 2}.
\label{iden}
\end{equation}
Hence from \eqref{fpsp}, $\srvq/\N \to 1$ as $N_t \to \infty$,
and substituting in \eqref{Nrp} implies that $\B \to 0$.

To determine the first-order rate at which $\B \to 0$,
we combine \eqref{nc3} and \eqref{iden} to write
\begin{equation}
\frac{\srvq}{\N} -1 = \frac{\log 2}{\mu \N} \T +  O(\T^2)
\label{srvqT}
\end{equation}
The behavior of $\srvq$ for small $\B$ (equivalently,
$\srvq/\N$ close to one) can be
determined by expanding $W(x)$ around $x = - \me^{-1}$.
Such an expansion is given in \cite{barry},
which we rewrite as
\begin{equation}
  \srvq = \N \left( 1 + \sqrt{\zb} +\frac{1}{3} \zb + \frac{11}{72}
  \zb \sqrt{\zb} + O(\zb^{5/2}) \right)
\label{szb}
\end{equation}
where $\zb = 2 (1 - 2^{-\B / \N}) = (2 \log 2) (\B/\N) + O(\B^2)$
for small $\B$. Hence we have
\begin{equation}
\frac{\srvq}{\N} -1 = \sqrt{\zb} + O(\zb) = 
\sqrt{ \frac{2 \log 2}{\N} } \sqrt{\B} + O(\B) .
\end{equation}
Combining this with \eqref{srvqT} gives
\begin{equation}
\sqrt{\B} = \frac1{\mu}\sqrt{\frac{\log 2}{2\N}} \T + O(\T^2) .
\end{equation}
and substituting for $\T$ from \eqref{Tulo}, we conclude that
the feedback overhead that maximizes the upper bound on 
achievable rate satisfies
\begin{equation}
\Bu = \frac{\Lb^2 \log 2}{2\mu^2 \N} \frac1{\log^2 N_t} 
+ O \left( \frac1{\log^4 N_t} \right)
\label{Bulo}
\end{equation}
Substituting for the optimized $\Tu$ and $\Bu$ in $\mC_u$ gives
\begin{equation}
  \mC_u^o - \log( \rho N_t) + \log \log N_t \longrightarrow \log
  \left( \frac{\rho \Lb \N}{\me (\rho + 1)} \right) .
\end{equation}

We can apply the same techniques to the lower bound on achievable rate
to determine the behavior of the optimal parameters.  (Here we must
show that $\cNt$ in \eqref{cNt} tends to zero uniformly over all $\T$
and $\B$.)  The training and feedback overhead that maximize the lower
bound on achievable rate satisfy
\begin{gather}
   \T^o_l \log N_t \longrightarrow \Lb \\ 
   \B^o_l \log^2 N_t  \longrightarrow  \frac{\Lb^2 \log 2}{2 \mu^2 \N}
\end{gather}
and substituting into the expression for $\mC_l^o$ gives
\begin{equation}
  \mC_l^o - \log( \rho N_t) + \log \log N_t \longrightarrow \log
  \left( \frac{\rho \Lb \N}{\me (\rho + 1)}\right) - \log(1 + \rho) .
\end{equation}
Since the lower and upper bounds grow at the same rate, this
establishes the theorem.

\bibliographystyle{IEEEtran}
\bibliography{IEEEabrv,train}

%%%% BIO'S %%%%%%%%%%%%%%%%%%%%%%%%%%

\begin{IEEEbiographynophoto}{Wiroonsak Santipach}
(S'00-M'06) received the B.S. ({\em summa cum laude}), M.S., and
Ph.D. degrees all in electrical engineering from Northwestern
University, Illinois, USA in 2000, 2001, and 2006, respectively.

He is currently an assistant professor at the Department of Electrical
Engineering, Faculty of Engineering, Kasetsart University in Bangkok,
Thailand.  His research interests are in signal processing for
wireless systems such as CDMA, MIMO, and OFDM.
\end{IEEEbiographynophoto}

\begin{IEEEbiographynophoto}{Michael L. Honig}
(S'80-M'81-SM'92-F'97) received the B.S. degree in electrical
  engineering from Stanford University in 1977, and the M.S. and
  Ph.D. degrees in electrical engineering from the University of
  California, Berkeley, in 1978 and 1981, respectively. He
  subsequently joined Bell Laboratories in Holmdel, NJ, where he
  worked on local area networks and voiceband data transmission. In
  1983 he joined the Systems Principles Research Division at Bellcore,
  where he worked on Digital Subscriber Lines and wireless
  communications.  Since the Fall of 1994, he has been with
  Northwestern University where he is a Professor in the Electrical
  Engineering and Computer Science Department.  He has held visiting
  scholar positions at the Technical University of Munich, Princeton
  University, the University of California, Berkeley, Naval Research
  Laboratory (San Diego), and the University of Sydney.  He has also
  worked as a free-lance trombonist.

Dr. Honig has served as an editor for the IEEE Transactions on
Information Theory (1998-2000), the IEEE Transactions on
Communications (1990-1995), and was a guest editor for the European
Transactions on Telecommunications and Wireless Personal
Communications. He has also served as a member of the Digital Signal
Processing Technical Committee for the IEEE Signal Processing Society,
and as a member of the Board of Governors for the Information Theory
Society (1997-2002).  He is the recipient of a Humboldt research award
for senior U.S. scientists, and the co-recipient of the 2002 IEEE
Communications Society and Information Theory Society Joint Paper
Award and the 2010 IEEE Marconi prize paper award.
\end{IEEEbiographynophoto}

\end{document}